\documentclass{aa501}
\usepackage{psfig,longtable,lscape}
\usepackage{graphicx}
\usepackage{rotating}

\begin{document}

\title{ADONIS observations of hard X-ray emitting late B-type stars
in Lindroos systems
\thanks{Based on observations collected at the European
Southern Observatory, La Silla, under project 65.H-0568(A)}}

\author{N. Hu\'elamo\inst{1}
\and W. Brandner\inst{2,3}
\and A.G.A. Brown\inst{3}
\and R. Neuh\"auser\inst{1,2}
\and H. Zinnecker\inst{4}}

\institute{Max-Planck-Institut f\"ur extraterrestrische Physik,
 Giessenbachstrasse 1, D-85741 Garching, Germany
\and University of Hawaii, Institut for Astronomy, 2680 Woodlawn Dr.,
   Honolulu, HI 96822, USA
\and European Southern Observatory, Karl Schwarzschildstrasse 2, 
   D-85748 Garching, Germany
\and Astrophysikalisches Institut Potsdam, An der Sternwarte 16,
  D-14482 Potsdam, Germany\\}

\offprints{N. Hu\'elamo, huelamo@mpe.mpg.de}

\date{Received/Accepted}

\abstract{We present adaptive optics $JHK_S$ imaging observations of three
main-sequence late B-type stars listed in the Lindroos Catalogue:
HD\,123445, HD\,127971 and HD\,129791. Given their spectral types,
these stars should not be X-ray emitters. However, they have been
detected by {\em ROSAT} and their X-ray emission has been attributed
to possible unresolved late-type companions. We have carried out
near-IR observations with ADONIS at the ESO 3.6m but have not detected
any late-type companions close to HD\,127971 and HD\,129791.
This result leads us to conclude that either (i)
they are spectroscopic binaries with unresolved low-mass 
companions, or (ii) they are intrinsic X-ray emitters.  While the
former case would be consistent with the reported high
multiplicity of early-type (A and B) stars, the latter would yield a
revision of stellar activity theories which do not predict X-ray
emission from these stars. On the other hand, HD\,123445 does indeed
show visual companions, namely an apparent subarcsecond faint
($K_s\sim10$) binary system at a projected separation of 5\arcsec~from
the late-B type star. The $JHK_S$ magnitudes and colors of the
components are consistent with (i) a pair of Pre Main Sequence (PMS)
K-type stars at 140\,pc (i.e. possible members of the Upper Centaurus
Lupus association), (ii) a pair of Main Sequence M-type stars at
60\,pc and (iii) a pair of K-type giants at 2.6\,kpc. While in the
first case the reported X-ray emission can be ascribed to the new
objects, in the second and third case it cannot, and we have to assume
the late B-type star to be either a spectroscopic binary itself or a
single star with intrinsic X-ray emission. Spectroscopy is required to
confirm the possible PMS nature of the new binary and {\em
Chandra} X-ray high spatial resolution (astrometric) imaging
observations are required to definitely determine the source of the
X-ray emission. If the B9 star results to be the X-ray emitter,
near-IR spectroscopy can be used to investigate the presence of a T
Tauri like spectroscopic companions. }
\maketitle

\section{Introduction}
 Lindroos systems are defined as visual binaries mainly comprised of
 Main Sequence (MS) early-type primaries and later-type secondaries
 (Lindroos 1985). The ages of the primaries have been derived trough
 photometric and spectroscopic observations, showing values of the
 order of $\sim$10$^7$ yr. Because this age is comparable to the
 contraction time scale of late-type stars to the MS,
 if the systems are bound the secondaries can be Pre-Main Sequence
 (PMS) stars still
 contracting to the Zero Age Main Sequence (ZAMS), that is, 
 T Tauri (TTS) or Post-T Tauri stars (PTTS).

Pallavicini et al. (1992) and Mart\'{\i}n et al. (1992) have reported
the presence of youth (Li {\sc I} absorption line) and activity
(H$\alpha$ emission line) indicators among the late-type
secondaries. Because the X-ray emission is another indicator of
youth/activity among late-type stars (Neuh\"auser et al. 1995), a
study of the X-ray properties of the Lindroos secondaries has been
undertaken to confirm their suspected PMS nature.
The X-ray emission from Lindroos systems has been studied by Hu\'elamo
et al. (2000), as an extension of the work by Schmitt et al. (1993).
The study was based on {\em ROSAT} High Resolution Imager (HRI) data,
which was the only instrument on board the {\em ROSAT} satellite
capable of resolving most of the systems.  The analysis of the
HRI X-ray data from 22 binary systems shows that 16 of the
resolved late-type secondaries are X-ray emitters, with X-ray luminosities 
and X-ray spectral energy distributions
comparable to those from PMS late-type stars (Neuh\"auser et al.  1995).

Surprisingly, the X-ray detection of several late B-type stars in
Lindroos systems has been also reported.  Theoretically, late B-type
and early A-type stars are not expected to be X-ray emitters. Unlike O
and early B-type stars they do not show strong winds, so the X-rays
cannot be produced by energetic shocks due to instabilities in the
radiatively driven winds (Lucy \& White 1980, Owocki 1988) and, in
contrast to late-type stars, they do not have a convective zone able
to power a corona.  Nevertheless, the X-ray emission from late-B and
early-A type stars as seen by Einstein and ROSAT satellites has
been reported by several authors (Caillault \& Zoonematkermani 1989,
Caillault et al. 1994, Bergh\"ofer \& Schmitt 1994, Simon et al. 1995,
Bergh\"ofer et al. 1997, Panzera et al. 1999). The most widely
accepted explanation for their X-ray emission is the presence of
unresolved late-type companions responsible of the X-ray detections.
However, the moderate spatial resolution of the Einstein and
ROSAT X-ray detectors has not allowed to probe this hypothesis, being
the intrinsic X-ray emission of late B-type stars still an open
question.

In the case of early-type stars in Lindroos systems, Hu\'elamo et
al. (2000) have found that 3 (out of 5) late B-type primaries show
X-ray properties very different in comparison with similar MS late
B-type stars.  Firstly, they show large X-ray luminosities comparable
to those from PMS late-type stars.  Secondly, the spectral
distribution of their X-ray emission is similar to that from TTS in
star forming regions, that is, they are hard X-ray emitters.  The fact
that these X-ray properties are also similar to those from the
Lindroos PMS late-type secondaries, has strengthened the hypothesis of
unresolved PMS late-type companions to these 3 B-type primaries.

In order to check this hypothesis we have carried
out diffraction limited IR observations of the 3 late B-type stars
in Lindroos systems with a higher probability 
of showing close companions. Our main aim is to check
if their X-ray emission can be ascribed to unknown late-type
stars, previously unresolved. If so, this result would strengthen
the idea that late B-type stars are not intrinsic X-ray emitters. If
not, it would suggest these late B-type stars are (i) spectroscopic
binaries with even closer late-type companions or (ii) intrinsic X-ray
sources. Note that an extended work on adaptive optics (AO) 
observations of  X-ray emitting late B-type stars 
has been recently carried out by Hubrig et
al. (2001) on different targets.

The main properties of the three late B-type stars are described in
Section 2, while the details of the data reduction are provided in
Section 3. The results for each source are discussed in Section 4. The
main conclusions are summarized in Section 5.

\begin{table*}
\caption{Main properties of the three Lindroos late B-type stars observed with ADONIS }
\begin{flushleft}
\begin{tabular}{ll@{\hspace{3mm}}c@{\hspace{3mm}}c@{\hspace{3mm}}c@{\hspace{3mm}}l@{\hspace{2mm}}c@{\hspace{2mm}}ccl}\hline\noalign{\smallskip}
{\bf Star} & {\bf Sp. T.}  &  {\bf V} & {\bf A$\bf_v$}$^1$ & {\bf Mass}$^1$ & {\bf Distance}$^2$ & {\bf Membership}$^3$ &
 \mbox{\boldmath $\bf\mu_{\alpha}^*$,$\bf\mu_{\delta}$} $^4$ & {\bf L$\bf_x$} $^5$ & {\bf HR}$^5$ \\
 & & (mag)  & (mag) & (M$_{\odot}$) & (pc) &  & (mas)& (10$^{30}$ erg/s)&  \\
\noalign{\smallskip}
\hline
\noalign{\smallskip}
HD\,123445  & B9   & 6.19 & 0.12 & 3.0 & 218$\pm$37 &Upper Cen Lup & -19.3$\pm$0.8,-17.4$\pm$0.6 & 3.47$\pm$0.51 & 0.19$\pm$0.13 \\
HD\,127971  & B7   & 5.89 & 0.08 & 3.5 & 109$\pm$9 & & -21.6$\pm$0.8,-23.2$\pm$0.7 & 0.55$\pm$0.08 & 0.49$\pm$0.13\\
HD\,129791  & B9.5 & 6.94 & 0.26 & 2.5 & 129$\pm$16  & Upper Cen Lup & -26.1$\pm$1.1,-20.8$\pm$0.8& 3.80$\pm$0.29 &
 0.56$\pm$0.07\\ \hline
\end{tabular}
\end{flushleft}
{\bf Notes to the table:} $^1$ Extracted from Lindroos (1985);
$^2$ Derived from Hipparcos parallaxes; $^3$ de Zeeuw et al. (1999);
$^4$ Hipparcos proper motions;
$^5$ Data from Hu\'elamo et al. (2000); the X-ray 
luminosities are derived using Hipparcos distances.

\end{table*}
\begin{figure*}
   \resizebox{18cm}{!}{\includegraphics{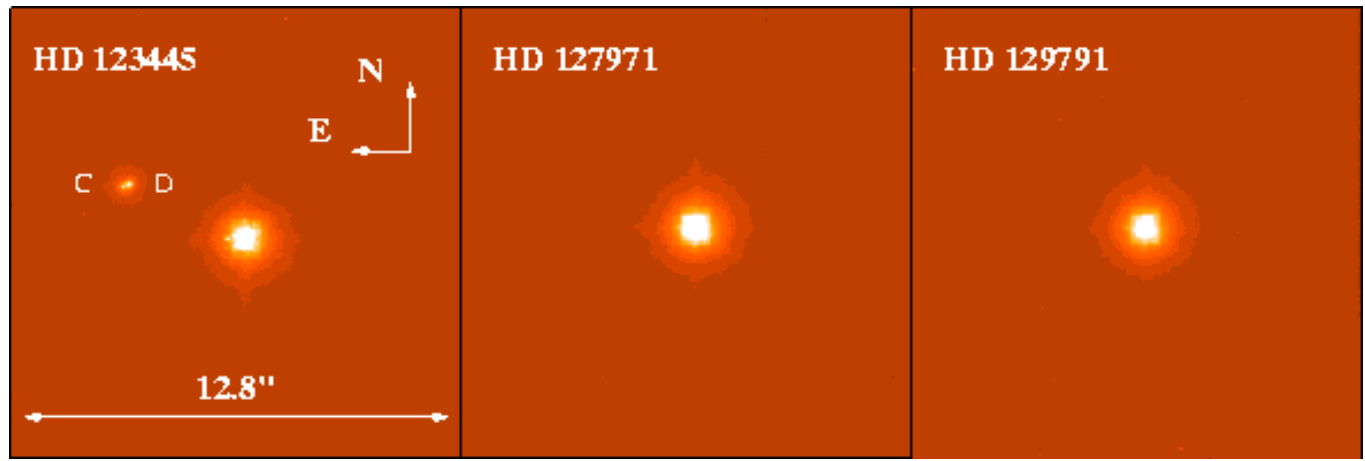}}
   \caption{{\em K$_S$}-band images of the three stars 
    observed with ADONIS. The
    whole field of the SHARP II camera (12.8\arcsec~x 12.8\arcsec) is shown.
    As can be seen, only in the case of HD\,123445 two new objects are
    detected at $\sim$ 5\arcsec~NE from the central star (perhaps a true
    0.25\arcsec binary system in the UCL association). In the case of
    HD\,129791 and HD\,127971 no extra sources are detected.}
\end{figure*}

\section{X-ray properties of the late B-type stars under study}

The three stars under study are included in the so-called Lindroos catalogue
(Lindroos 1985), and have late-type visual companions. All of them have been
detected and resolved by the {\em ROSAT} High Resolution Imager (HRI), showing
unusual X-ray properties in comparison with similar MS early-type stars.

As shown in Hu\'elamo et al. (2000), MS early-type stars in Lindroos systems
show X-ray luminosities which decrease with later spectral types.  While early
B-type stars show X-ray emissions comparable to that from PMS objects, late
B-type and A-type stars show X-ray luminosities of $\log\,L_x(erg/s)\sim 28$.
The three late B-type stars under study in the present paper show X-ray
luminosities larger than $\log\,L_x(erg/s) >29.5$, that is, clearly comparable
to that from TTS.

Spectral information of the X-ray emission can be also derived through
the so-called hardness ratio {\em (HR)}. The HRI pulse height distribution
can be used to compute a two-band (soft and hard) hardness ratio (see
Hu\'elamo et al. 2000 for details).  The derived {\em HR} for the sample of
Lindroos primaries shows that they are generally soft X-ray emitters
with $HR<0$, while the PMS late-type secondaries are hard X-ray
sources with $HR>0$. In the case of the three late B-type stars with
X-ray emission, their {\em HR} is similar to that from young late-type stars
(see Table 1).

Also note that a mean ratio of $\log L_x/L_{bol}\sim-6$ has been
derived for late B-type stars observed in the {\em RASS} (Bergh\"ofer
et al. 1997).  This value is two orders of magnitude larger than the
$\log L_x/L_{bol}\sim-8$ value derived by Cassinelli et al. (1994) for
stars with spectral types later than $\sim$B2. Bergh\"ofer et
al. (1997) have attributed such a large ratio to possible late-type
companions not resolvable in the {\em RASS} and responsible of the
X-ray detection.  The three Lindroos late-B type stars under study
show $\log L_x/L_{bol}$ ratios between -4.5 and -6, so they show
large values that could certainly be related with the presence of
X-ray emitting unresolved companions.  

Therefore, our AO analysis is focused on the three late B-type
Lindroos primaries stars resolved by the {\em ROSAT} HRI showing high
X-ray luminosities and {\em HR} typical of late-type stars, that is,
we will study those late B-type stars with a higher probability of
having an unresolved PMS late-type companion.  Table 1 shows the
stellar data for the three selected sources.  Columns 1 and 2 provide
the HD number and the spectral type of the stars. Columns 3, 4 and 5
show the visual magnitude, the visual extinction to the source and the
stellar mass (according to Lindroos, 1985), respectively.  The
distance to the source and its proper motion, both derived from
Hipparcos data, are provided in columns 6 and 7.  The OB-association
to which the stars probably belong (de Zeeuw et al. 1999) is
given in Column 8.  Note that in the case of HD\,127971, we have
adopted the result by de Zeeuw et al. (1999) who do not find clear
evidence of membership (to the Upper Centaurus Lupus association) of
this star based on Hipparcos data.  Finally, the derived X-ray
luminosity of the sources together with their {\em HR} (Hu\'elamo et
al. 2000) are provided in Columns 9 and 10.

\section{The ADONIS data}

The IR data of the three late B-type stars were obtained with the Adaptive
Optics Near Infrared System (ADONIS) plus the System for High Angular
Resolution Pictures (SHARP) II camera at the 3.6m telescope at La Silla
Observatory. The main advantage of the AO system is that provides diffraction
limited resolution in the core of the Point Spread Function (PSF)
(i.e. 0.15\arcsec~in the K-band).  The observations were carried out on the
2$^{nd}$ June 2000, in the course of the programme 65.H-0658(A).

The three targets are brighter than 8 mags in the optical (see Table 1) so the
Reticon wave front sensor was used to perform the atmospheric correction. The
observations were carried out in the $JHK_S$ filters. For each source,
120 individual frames (30 frames each at 4 different positions of the
detector) were taken in the three filters. The individual exposure times range
between 0.3 and 0.8 sec, and were chosen so as to avoid saturating the
detector.  Several PSF-calibrator and standard stars were also observed
during the night.

Note that the three 
late B-type stars under study have known visual companions
located at more than 15\arcsec~(see Lindroos 1986).  Because the field of view
(FOV) of the SHARP II camera is 12.8\arcsec~x12.8\arcsec (with a plate scale
of 0.05\arcsec/pixel), these companions are located outside the frames.
\begin{table*}
\caption{Observational data of the 3 stars under study.}
\begin{flushleft}
\begin{tabular}{llllllcccc}
\noalign{\smallskip}
\hline
\noalign{\smallskip}
{\bf HD} & {\bf Band} &\multicolumn{3}{c}{{\bf Strehl ratio} ($\%$)} & {\bf N$^o$ frames} & {\bf Exp. time} & \multicolumn{3}{c}{\bf Limiting magnitudes}\\
 &    & Mean   & Min.$^1$ & Max. & selected &  & 0.5\arcsec & 1\arcsec & 2\arcsec \\
 & & & & & & (sec) & (mag) & (mag) & (mag)\\
\noalign{\smallskip}
\hline
\noalign{\smallskip}
123445     & {\em J}     & 6  &  5  & 7.2& 38 & 0.5 & 14.2 & 16.2 & 19.5\\
           & {\em H}     & 14 & 10  & 16 & 73 & 0.3 & 13.1 & 16.5 & 19.4\\
           & {\em K$_S$} & 28 & 25  & 32 & 61 & 0.5 & 15.1 & 16.3 & 18.9\\ 
\noalign{\smallskip}
127971     & {\em J}     & 1.4 & 1  & 2.2& 64 & 0.6 & 14.5 & 16.2 & 18.6\\
           & {\em H}     & 5   & 3  & 8  & 34 & 0.6 & 14.3 & 15.8 & 18.6\\
           & {\em K$_S$} & 45  & 30 & 52 & 100& 0.4 & 13.8 & 16.6 & 19.1\\ 
\noalign{\smallskip}
129791     & {\em J}     & 3   & 2.5& 4  & 21 & 0.8 & 14.7 & 16.9 & 19.5\\
           & {\em H}     & 7   & 5  & 12 & 51 & 0.8 & 15.0 & 16.9 & 19.8\\
          & {\em K$_S$}  & 20  & 15 & 24 & 56 & 0.6 & 15.1 & 17.4 & 19.8\\ \hline
\end{tabular}
\end{flushleft}
{\bf Notes to the Table:} {$^1$}Minimum Strehl ratio
selected to improve the quality of the final images.
\end{table*}

The data reduction was performed using the Eclipse (Devillard, 1997) and the
IRAF packages. All the images are sky-subtracted, dark and flat-field
corrected. A bad-pixel map correction was also applied.  The final images are
the result of averaging the individual frames with higher quality, ie. with
larger Strehl ratio (SR) and smaller FWHM. This was necessary due to the rapid
seeing variations during the night (from 0.7\arcsec~to 1.6\arcsec).

We have computed the limiting magnitudes of the final frames in each filter,
assuming a 3$\sigma$ level over the background as a detection.  We have
derived these magnitudes at three different positions from the central star,
namely 0.5\arcsec, 1\arcsec and 2\arcsec, in order to gain a better
understanding of the detection limits. Note that the resulting magnitudes are
dependent of several factors: the brightness of the central object, the Strehl
ratio, the number of averaged individual images and the individual exposure
time which makes the readout noise (RON) an important limitation.

 Kirkpatrick \& McCarthy (1994) 
 provide absolute near-IR magnitudes of M-dwarfs down to 
 the substellar limit (M7 to M9 dwarfs).  Using these values we have
 derived the expected near-IR magnitudes of late M-dwarfs at the
 distances of our three targets (that is, 218, 109 and 129\,pc). 
 These data together with the limiting magnitudes
 provided in Table 2 show that we could detect 
 companions around the three sources down to a spectral type 
 of M9 at projected separations larger than 1\arcsec.

The observational data are summarized in Table 2. The name of the star and the
filter are provided in Columns 1 and 2. The minimum SR selected for each of
the final frames, together with the maximum and mean values, are given in
Columns 3, 4 and 5. The total number of frames (out of a total of 120) used
for the final averaging and their individual exposure times are given in
columns 6 and 7.  The limiting magnitudes for each of the final frames at
0.5\arcsec, 1\arcsec and 2\arcsec~from the late B-type star are provided in
the last columns of Table 2.

\section{Analysis of the data}

 We have derived the near-IR magnitudes of all the detected objects
 after carrying out aperture photometry of the sources. The selected
 aperture radius for all the targets is 50 pixels (2.5\arcsec).  The
 zero-points in the three filters were derived using standard
 stars observed during the night. We obtained the following values:
 21.5, 21.9 and 22.2 {\rm mags}, in the {\em K$_S$}, {\em H} and {\em
 J}-band, respectively.  The mean extinction coefficients at La Silla
 for the three filters were used. The main uncertainties of the IR
 magnitudes come from the errors associated to the zero points and the
 extinction coefficients in each band, resulting in a typical error of
 $\pm$0.1 mags in the three filters.

 The analysis of the final images shows that 
 in the case of HD\,123445 two new sources have been detected at
 5\arcsec~NE from the B9 star.  As it is seen in Figure 1, the two
 objects are very close to each other (0.26\arcsec$\pm$0.01), so their
 IR magnitudes were determined through multi-aperture photometry.  In
 the case of HD\,127971 and HD\,129791 no extra objects were detected
 in the field.

 The derived near-IR magnitudes for the 5 detected objects are listed
 in Table 3. In the case of the late B-type stars, they are in good
 agreement with those reported by Lindroos (1983).  Also the derived
 IR photometric distances are in good agreement with the Hipparcos
 distances, allowing for an error of one subtype in the spectral
 types.

 The {\em J-H} and {\em H-K$_S$} colors have been computed for the 5
 objects.  Those are represented in the color-color diagram shown in
 Figure 2.  We will discuss the results for each star in the following
 subsections.

\subsection{HD\,123445}

 HD\,123445 is a B9 type star with a known visual companion, {\it
 B}, at a separation of 28.6\arcsec. In our observations, we have
 detected two new sources in the field of the B9 star.  The two
 objects ({\it C} and {\it D} in Figure 1) are located at
 4.58$\pm$0.01\arcsec~NE and 4.46$\pm$0.01\arcsec~NE from
 HD\,123445, respectively. Their position angles (PA), measured
 from the north to the east, are $ PA(C)=\,65\pm1^\circ$ and
 $PA(D)=\,64\pm1^\circ$. From the Hipparcos parallax, the distance to
 HD\,123445 is 218 \,pc. Hence, the projected separation of these
 stars with respect to the B9 primary is $\sim$1000 AU.  Note,
 however, that the uncertainty in the Hipparcos distance is large, so
 the star could be located much closer (181\, pc) or much further
 (255\,pc) in the association.  Given that the mean distance to the
 UCL association is 140\,pc with a depth of $\pm 50$\,pc (de Bruijne
 et al. 1999), it is most probable that HD\,123445 is at the back end
 of the association with a distance between 180 and 190\,pc.
 
  As a first step, we have studied the {\em ROSAT} HRI image of
  HD\,123445 to check if the position of the single X-ray detection
  is consistent with the optical position of the new
  binary.  Figure 3 shows the {\em ROSAT} HRI X-ray image of
  HD\,123445. The nominal positional error of the HRI is 5\arcsec,
  although a bore-sight correction as much as 10\arcsec~can be
  applied. Therefore, we have plotted the position of the X-ray
  detection enclosed by a circle of 10\arcsec~radius.  The Hipparcos
  optical position of HD\,123445 (open triangle) together with the
  optical position of the two new sources (open hexagon) are
  over-plotted.  As noted in Hu\'elamo et al. (2000), the single X-ray
  detection is displaced 7.7\arcsec~from the optical position of
  HD\,123445.  However, the new binary is certainly closer to the
  X-ray detection than the late B-type star, being located at
  4.2\arcsec~SW from the X-ray source.
\begin{table}
\caption{Near-IR photometric data of the 5 objects detected}
\begin{flushleft}
\begin{tabular}{lrrrcc}
\noalign{\smallskip}
\hline
\noalign{\smallskip}
{\bf Star} &  {\bf J} & {\bf H} &  {\bf K$\bf_S$} & {\bf J-H} & {\bf H-K$\bf_S$}\\
  & (mag) & (mag) & (mag) & (mag) & (mag) \\
\noalign{\smallskip}
\hline
\noalign{\smallskip}
HD\,123445       & 6.2  & 6.2  &  6.2 & 0.0 & 0.0 \\
Comp. {\it C}     & 10.8 & 10.1 & 9.9 & 0.7 & 0.2 \\
Comp. {\it D}     & 10.7 & 10.1 & 9.9 & 0.6 & 0.2 \\
HD\,127971       & 6.0 & 5.9 & 5.9 & 0.1 & 0.0 \\
HD\,129791       & 6.7 & 6.6 & 6.6 & 0.1 & 0.0 \\
\noalign{\smallskip}
\hline
\noalign{\smallskip}
\end{tabular}
\end{flushleft}
{\bf Note:} The errors in the magnitudes are $\pm$0.1 in the three bands
\end{table}
 \begin{figure} \vspace{-0.4cm}
  \resizebox{9cm}{!}{\includegraphics{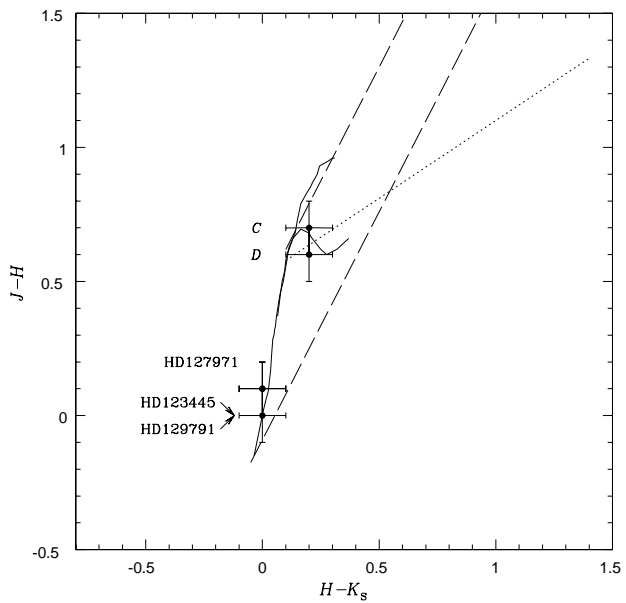}}
 \caption{Color-color diagram for the 3 stars under study.  We have
 represented by solid lines the unreddened main sequence and giant
 sequence from Bessell \& Brett (1988). The dashed lines represent the
 reddening vectors (Rieke \& Lebofsky 1985) while the dotted line
 represents the color-color relation followed by WTTS and CTTS in
 Taurus (Meyer et al. 1997). The three late B-type stars lie at the
 left bottom of the diagram. The two sources located at 5\arcsec~SE
 from HD\,123445 lie in the upper region of the diagram close to
 the K-giants and the MS M-type stars.  These sources also followed
 the color-color relation found for PMS late-type stars, CTTS
 and WTTS, in Taurus (Meyer et al. 1997) within the errors.}
 \end{figure}

 Apart from HD\,123445, there are no other bright X-ray sources in
 the FOV of the HRI, so it is not possible to measure relative
 positions through X-ray field objects, i.e. we could not do a proper
 bore-sight correction. Hence, we can firstly conclude that although
 the X-ray detection is closer and therefore possibly related to the
 new objects (C and D), it could also be ascribed to HD\,123445
 itself.  

 As a second step, we have analyzed the near-IR data from the detected binary.
 The IR magnitudes of HD\,123445 together with those from the two new
 objects are listed in Table 3. It can be seen that the two members of the new
 detected binary show very similar IR magnitudes and colors. When we study the
 position of the two faint objects in a color-color diagram (Fig. 2), we can
 see that within the errors both lie close to the region of the MS and PMS M
 and K-type stars. Their colors are also consistent with those from giant
 K-type stars. We will analyze each of these possibilities to study if any of
 them is in good agreement with the derived X-ray data:

 {\bf a)} Both objects show IR colors consistent with M0V to M5V
 stars. Therefore, we will assume an intermediate spectral type of M2
 for both sources.

 According to Kirkpatrick \& McCarthy (1994), the absolute H-magnitude
 of a M2V star is $M_H=6.21$. Using the derived H-magnitudes of our
 objects and the distance modulus relation, we can derive a
 photometric distance of 60\,pc. In this case, the objects would not
 be bound to HD\,123445.  If they were so close, they would not
 belong to any known star forming region, so we will assume them to be
 MS M-type stars but no PMS objects.  However, this assumption will be
 discussed later.

 Gliese \& Jahrei{\ss} (1991) and Henry et al. (1994) have identified
 a large sample of MS low-mass stars in the solar neighborhood
 ($d\,<\,25\,pc$). The X-ray emission of these nearby MS late-type
 stars has been studied by Schmitt et al. (1995), Schmitt (1997) and
 H\"unsch et al. (1999) based in {\em ROSAT} data.  The mean/median
 X-ray luminosity reported for MV type stars is $L_x(erg/s)
 \sim 5.5\times10^{28}/8.6\times10^{27})$, with 
 a large dispersion of $\sigma\sim 1.3\times10^{29}/2.2\times10^{28}$,
 respectively.

\begin{figure}
\begin{center}
 \resizebox{8.5cm}{!}{\includegraphics{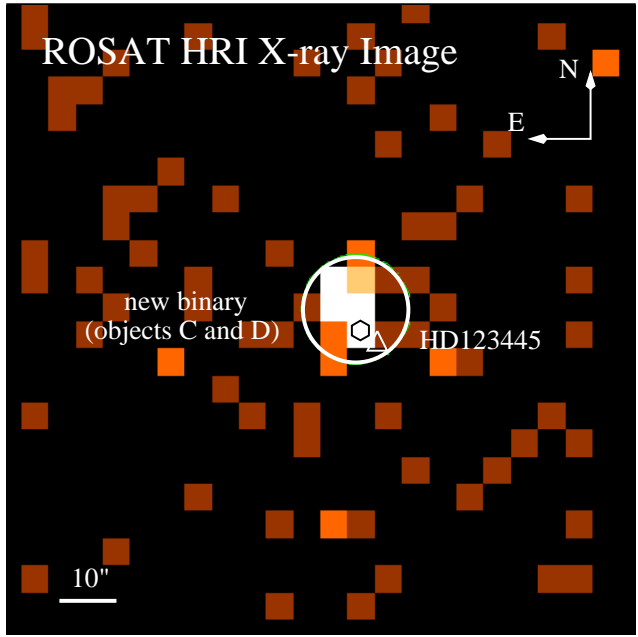}}
 \caption{{\em ROSAT} X-ray image of HD\,123445.  The X-ray source is
 located inside the white circle, whose radius is 10\arcsec, that is,
 it represents the maximum positional error of the {\em ROSAT} HRI.  The
 optical positions of HD\,123445 and the new binary ({\it C} and {\it D}
 objects) are represented by an open triangle and a hexagon,
 respectively. 
 As seen, the new subarcsecond binary lie closer to the
 X-ray detection than the late B-type star. However, both positions
 are within the HRI positional error, so the X-ray emission could be
 ascribed to any of them.}
\end{center}
\end{figure}
  
  Making use of the mean X-ray luminosity (in order to be conservative and
  to include the strongest X-ray emitters), we can derive the expected 
  {\em ROSAT} HRI count rate from a M-type star at 60\,pc. 
  The X-ray luminosity of a
  source is given by the expression:
  \begin{equation}
  L_x = \frac{4\pi d^2}{ECF}\times(rate)
  \end{equation}
  where {\it d} is the distance to
  the source, {\it ECF} the energy-conversion-factor and {\em rate}
  the X-ray counts per second coming from the source.
  The {\it ECF} depends on the detector response and the underlying
  model for the X-ray spectrum.  In our case, we have assumed a
  Raymond-Smith thermal spectrum with temperature $kT=1keV$ (typical
  for late-type stars), and a negligible extinction to the source
  ($lg\,N_H (cm^{-2})$ = 19.0). In accordance with the HRI response
  curve, the derived $ECF$ is $0.20\,10^{11}\, cts \cdot cm^{-2} \cdot
  erg^{-1}$.  Following eq. (1), the expected HRI count rate for a
  M-type star at 60\,pc is $rate(cts/s)=0.002$.  We have detected two
  objects in the near-IR images.  If we assume they emit the same
  amount of X-rays, the total expected count rate would be
  0.004$cts/s$.  However, when studying the HRI X-ray detection
  of HD\,123445, the derived count rate is $rate(cts/s)=0.012\pm0.002$, 
  which is larger than that expected from two MS M-type stars. Note that the
  X-ray emission from HD\,123445 was shown to be from a quiescent,
  non-flaring source (Hu\'elamo et al. 2000).

  The large scatter in the X-ray luminosities of MS
  M-type stars suggests very different properties among the sample,  
  and does not allow us to clearly conclude that the new
  objects are inconsistent with this spectral type. As a further test,
  we have made use of the empirical saturation limit of
  $\log(L_x/L_{bol})=-3$ for late-type stars to compare it with that
  derived for our stars. The bolometric luminosity for a MS M2-type star
  at 60 pc (using the bolometric correction from Schmidt-Kaler 1982),
  results in $L_{bol}(erg/s)= 1.42\times10^{32}$.  
  If we suppose that all the X-rays (0.012 cts/s) 
  are emitted only by one of the objects,
  and assuming an ECF as for late-type coronal sources ($0.20\,10^{11}\, 
  cts \cdot cm^{-2} \cdot erg^{-1}$)
  the derived $\log(L_x/L_{bol})$ ratio is larger than 
  the derived saturation limit of -3 for late-type stars.
  On the other hand, if we assume that each of the two objects emits 
  half of the X-rays, the derived X-ray luminosity 
  assuming the same ECF would be 
  $ L_x (erg/s)= 1.29\times10^{29}$.
  Then, the derived $\log(L_x/L_{bol})$ ratio for both objects is
  -3.04, that is, similar to the saturation limit of -3. 
  This implies that both objects must be very active M-type stars.  
  However, according to
  Schmitt et al. (1996) the X-ray emission from MS late-type stars in
  the solar neighborhood seems to be generally moderate.  In fact,
  most of the MS M-type nearby stars studied by H\"unsch et
  al. (1999) show negative {\em HR}, that is, they are soft X-ray
  sources mainly due to their lower levels of activity and because of
  negligible interstellar absorption at small distances. 
  Therefore, the probability of having two very active MS M-type
  stars at 60\,pc seems to be low according to these X-ray studies.
 
   Since young late-type objects in the solar neighborhood also seem
   to be abundant (Sterzik \& Schmitt 1997), we could also have
   assumed both objects to be PMS M-type stars.  Although there are
   only two M-type stars among the sample of most active nearby young
   stars, it seems that their X-ray emission is two orders of
   magnitude higher than in MS M-type stars. However, their {\em HR}
   is generally negative in part because of the negligible
   interstellar absorption.  This absorption is not significant at
   60\,pc.  In fact, Stelzer \& Neuh\"auser (2000) show that most of
   the young late-type stars which belong to the nearby Tucanae and TW
   Hya associations (d$\sim$50\,pc, age$\sim$ 10 Myr) show negative
   {\em HR}.  Therefore, even if the new objects were young M-type
   stars, this would not be consistent with our derived {\em HR},
   which is more common from young objects in star forming regions
   (Neuh\"auser et al. 1995).

  {\bf b)} We will now assume both objects to be K2-type stars. If we
  follow the same line of reasoning as before, and use the absolute visual
  magnitudes from Schmidt-Kaler (1982), we can derive a photometric
  distance of 137\,pc. This value is very close to
  the derived distance to the UCL association (140\,pc), so we
  will assume the new binary to be a member of this group.
  Note that, as in the previous case, both objects would not be bound
  to the late B-type star.
  
  According to de Geus et al. (1989) the estimated age of the UCL
  association is 13 Myr. That means that our two K-type stars could be
  young objects still contracting to the MS, that is, they might be T
  Tauri stars.  As can be seen from their IR colors, they do not show
  significant IR excesses. Hence, they could be considered as
  Weak-line T Tauri stars (WTTS), that is, young stars without
  optically thick accretion disks.  In fact, we show in Figure 2 that
  both objects follow the color-color relation 
  \begin{equation} 
  J-H =(0.58\pm0.11)\times(H-K) + 0.52\pm0.06 
  \end{equation} 
  found by Meyer et al. (1998) for WTTS and CTTS in Taurus, 
  within the uncertainties.

  The typical X-ray emission from K-type WTTS in Taurus is $\log\,L_x(erg/s)
  = 29.80\pm0.12$ (Stelzer \& Neuh\"auser 2001).  If we consider this
  luminosity, we can repeat the procedure described above assuming the
  new binary to be comprised of two K-type stars at 140\,pc emitting
  the same amount of X-rays. Hence, the derived count rate for the
  {\em ROSAT} HRI is $rate(cts/s)=0.007$ for one object, and $rate(cts/s)=
  0.014$ for two similar ones.  This value is certainly in very good
  agreement with the computed count rate for HD\,123445
  ($rate(cts/s)=0.012\pm0.002$).  Note that this result would imply
  the need of two objects at the same distance to produce the observed
  X-ray luminosity.

  Repeating the same procedure as in case (a) we have derived the
  bolometric luminosity of a K2-type star at 137\,pc, obtaining a
  value of $L_{bol}(erg/s)\sim9.8\times10^{32}$.  If we assume that
  just one of the sources emits all the X-rays, the derived X-ray
  luminosity at 137\,pc is $log L_x (erg/s) = 1.34\times10^{30}$.  In
  this case the $\log (L_x/L_{bol)}$ ratio is larger than the
  saturation limit of -3 for late-type stars.  However, if we assume
  that each of the sources emits half of the X-rays, the derived
  $\log(L_x/L_{bol)}$ ratio is equal to $7\times10^{-4}$, that is,
  close to the saturation limit.  If we take into account that the
  $\log (L_x/L_{bol})$ ratio of TTS in Taurus ranges between -3 and -5
  (Stelzer \& Neuh\"auser 2001), the result is consistent with both
  stars being very active PMS objects.
 
  Concerning the {\em HR}, most of the WTTS listed in Stelzer \&
  Neuh\"auser show positives values, meaning that they are mainly hard
  X-ray sources.  Therefore, both the X-ray luminosity and the {\em
  HR} can be reproduced assuming that the X-ray emission from
  HD\,123445 originates by two PMS K-type stars at 140\,pc.

 {\bf c)} The final possibility under study is that both objects are
 background evolved stars, given that their color indexes are also
 compatible with K-type giants (from K0 to K4). A K2 giant has typical
 colors of $V-K= 2.70$, $J-H = 0.63$ and $H-K=0.12$ and a visual
 absolute magnitude of $M_v= +0.5$ (Schmidt-Kaler, 1982). Using these
 data and the distance modulus relation we can derive a photometric
 distance of 2.6\,kpc to both sources.

 Late-type giants are easily detected in IR wavelengths given that
 their effective temperature (4000K) and their extended atmospheres
 makes them strong emitters in this spectral range (see Alves et
 al. 1998 for further discussion). The Galactic coordinates of
 HD\,123445 are $\ell(^\circ)$=317.52 and $b(^\circ)$= 17.20, so we
 are studying a region relatively close to the galactic bulge,
 increasing the probability of detection of red giant stars.

 On the other hand, late-type giant stars are known to be X-ray
 sources (Ayres et al. 1995, H\"unsch et al. 1998).  However, given
 their typical X-ray fluxes they could not have been detected at a
 distance of 2.6\,kpc by {\em ROSAT}. In this case, we would have to
 relate the reported X-ray emission to HD\,123445 itself, implying
 that it would probably be a spectroscopic binary or an intrinsic
 X-ray emitter.

 If we assume HD\,123445 to be a spectroscopic binary, we can roughly
 estimate the spectral type of the supposed companion making use of
 the derived X-ray luminosity. As in Section 4, we will assume a
 Raymond-Smith thermal spectrum with a temperature of 1keV.  The $ECF$
 is derived considering the $A_V$ listed in Table 1.  The result is
 that a late-type star at the same distance as HD\,123445 would show
 an X-ray luminosity of $\log\,L_x(erg/s)=30.60$.  This value agrees
 with that reported for PMS G-type stars in Taurus (Stelzer \&
 Neuh\"auser 2001).  Note, however, that in the case of HD\,123445
 there are no indications of a close companion in accordance with the
 Hipparcos Catalogue.

 After the analysis of the three possibilities, it seems that the case
 of two PMS K-type stars is in good agreement with the derived X-ray
 data. However, we note that spectroscopy is required before drawing
 final conclusions about the nature of these two sources. In
 particular, the detection of the Li I (6708\AA) absorption line would
 be a good indication of their PMS nature.  Another important clue
 about the true distance of the subarcsecond pair could be derived
 from its orbital motion.  Note that if the two objects are a true UCL
 binary, its orbital period would be $\sim$ 70 yr ( and $\sim$22 yr
 for a foreground binary with two M-type stars at 60 \,pc). Therefore,
 we could easily detect its orbital motion carrying out new IR
 observations in a few years.

\subsection{HD\,127971 and HD\,129791}

 In the case of HD\,127971 and HD\,129791 no extra sources have been
 detected in their surroundings (neither late-type stars nor
 substellar companions). This result implies that either (i) they are
 spectroscopic binaries with unresolved late-type stars or (ii) binary
 systems with substellar companions at a distances smaller than
 1\arcsec~or (iii) they are single B-type stars with intrinsic X-ray
 emission.

 Note that both stars are not resolved into multiple systems by
 Hipparcos. However, the Hipparcos Catalogue provides an acceleration
 solution for HD\,129791 which suggests the presence of an unresolved
 lower-mass companion.  In fact, the possibility of having two
 spectroscopic binaries is supported by several studies showing that
 the multiplicity of OB stars is rather high (i.e. Abt 1983, Abt et
 al. 1990, Morrell \& Levato 1991, Quist \& Lindegren 2000).

 We have roughly estimated the spectral types of the possible
 unresolved companions to these B-type stars, following the same
 procedure described above.  In the case of HD\,127971, the derived
 X-ray luminosity for a late-type star at the same distance would be
 $\log\,L_x(erg/s)=29.80$, which is consistent with a young K-type
 star (Stelzer \& Neuh\"auser 2001). For HD\,129791, the derived X-ray
 luminosity is $\log\,L_x(erg/s)=30.60$, which in good agreement with
 a PMS G-type star (as in the case of HD\,123445).
 This X-ray luminosity is
 larger than that reported for HD\,129791\,B
 \footnote{Note that 
 there is a typo in the X-ray luminosity
 of HD129791\,B in Table 5 of Hu\'elamo et al. (2000), being 
 the correct value $L_x = (6.64\pm1.33)\,10^{29}\,erg/s$}  
 (a PMS K5-type star probably bound to the B9 
 primary star), which would be consistent with their 
 difference in spectral types. 
   
 Note that if the lack of intrinsic X-ray emission from late B-type and early
 A-type stars could be demonstrated, then X-ray observations will be a suitable
 method for deriving the frequency of spectroscopic late-type companions among
 these intermediate-mass stars.

\section{Results and conclusions}

 The ADONIS IR {\em JHK$_S$} observations of 3 Lindroos late B-type 
 stars have
 allowed us to study the presence/absence of close visual companions around
 them which could possibly explain their X-ray emission. Our main conclusions
 can be summarized as follows:

{\bf 1.} Neither HD\,127971 nor HD\,129791 show objects around them in the AO
images, implying that close visual late-type companions are not the
explanation for their X-ray emission.  Further investigation (near-infrared
spectroscopy) is required in order to find out if these two stars are
spectroscopic binaries with unresolved late-type companions or if they are
intrinsic X-ray sources.

However, in the case of HD\,123445 two new objects are detected close to the
late B-type star. Although the nature of both sources can only be determined
through spectroscopic measurements, near IR magnitudes and colors show that
they are most probably PMS K-type stars at a distance of 140 \,pc (projected
towards but not physically bound to HD\,123445). In that case, the derived
photometric distance is in good agreement with that from the UCL association
to which HD\,123445 also belongs.  Moreover, the analysis of the X-ray data
suggests that they could certainly be responsible of the hard X-ray emission
associated with HD\,123445.

{\bf 2.} The X-ray data from HD\,127971 and HD\,129791 have allowed us to
roughly estimate the spectral types of possible unresolved companions. In both
cases, a young late-type star (a G-type and a K-type respectively) could be
responsible for the X-ray emission. With these late-type companions (masses
equal to or less than 1\,$M_{\odot}$), the mass ratio of the spectroscopic
binaries would be smaller than 1.

If it can be established that early A and late-type B stars lack
intrinsic X-ray emission, then a fruitful means of systematically
searching for lower-mass binary companions around this class of
primary stars (of order 3-4 M$_{\odot}$) would be to carry out
snapshot {\em Chandra} observations of a sample of early A and
late-type B-stars. The main goal would be to determine their binary
frequency and their mass ratio distribution, and to see if their
binary properties differ from those of more massive or lower mass
primaries, in particular solar-type primaries (Duquennoy and Mayor
1991).

\acknowledgements

We are grateful to F. Eisenhauer, L. Close M. Petr and B. Brandl for
their assistance during the data reduction.  Special thanks to
J. Alves, B. K\"onig and G. Wuchterl for their useful comments.
The {\em ROSAT} project is supported by the Max-Planck-Society
and the German Goverment (DRL/BMBF). WB acknowledges support
by NSF and NASA. RN wishes to acknowledge
financial support from the Bundenministerium f\"ur Bildung und Forschung
through the Deutsche Zentrum f\"ur Luft- und Raumfahrt e.V. (DLR) under
grant number 50 OR 0003.


\begin{thebibliography}{}


\bibitem{} Abt H.A., 1983, ARAA 21, 343

\bibitem{} Abt H.A., G\'omez A.E. and Levy S.G., 1990, ApJS, 74, 551

\bibitem{} Alves J., Lada C.J., Lada E., Kenyon S.J. and
     Phelps R., 1998, ApJ 506, 292

\bibitem{} Ayres T.R., Fleming T.A., Simon T., Haisch B.M. et al., 1995,
      ApJS 96, 223

\bibitem{} Bergh\"ofer T.W., Schmitt J.H.M.M., 1994,
      A\&A 292, L5


\bibitem{} Bergh\"ofer T.W., Schmitt J.H.M.M., Danner R.,
      Cassinelli J.P., 1997, A\&A 322, 167 

\bibitem{} Bessell M. and Brett J., 1988, PASP 100, 1134

\bibitem{} Caillault J.P. and Zoonematkermani S., 1989, ApJ 338, L57

\bibitem{} Caillault J.P., Gagne M. and Stauffer J.R., 1994, ApJ 432, 386

\bibitem{} Cassinelli J.P, Cohen D.H., MacFarlane J.J., Sanders W.T. and  
 Welsch B.Y, 1994, ApJ 421, 705

\bibitem{} de Bruijne, J.H.J., 1999, MNRAS 310, 585

\bibitem{} de Geus E.J., de Zeeuw P.T. and Lub J., 1989, A\&A 216, 44

\bibitem{} de Zeeuw P.T., Hoogerwerf R., de  Bruijne J.H.J, Brown A.G.A.
  and Blaauw A., 1999, AJ 117, 354

\bibitem{} Devillard N., 1997, 'The eclipse software', The messenger N$^o$87

\bibitem{} Duquennoy and Mayor M., 1991, A\&A 248, 485


\bibitem{} Gliese W. and Jahrei{\ss} H., 1991, {\it Preliminary version
 of the third Catalogue of Nearby Stars}, Brotzmann L.E. and Gesser S.E. eds.,
 The Astronomical data Center CD-ROM: selected Astronomical Catalogues, Vol. 1;
 NASA Astronomical Data Center, Goddard Space Flight Center, Greenbelt, MD.

\bibitem{} Henry T.J., Kirkpatrick J.D. and Simons D.A., 1994,
   AJ 108, 1437

\bibitem{} Hubrig S., Le Mignant D., North P. and Krautter J., 2001, 
            A\&A, in press

\bibitem{} Hu\'elamo N., Neuh\"auser R., Stelzer B., Supper R.
 and Zinnecker H., 2000, A\&A 359, 227

\bibitem{} H\"unsch M. , Schmitt J.H.M.M. and Voges W., 1998,
    A\&AS 127, 251

\bibitem{} H\"unsch M., Schmitt J.H.M.M., Sterzik M.F. and Voges W.,
    1999, A\&AS 135, 319

\bibitem{} Kirkpatrick J.D. and McCarthy D.W., 1994, AJ 107, 333

\bibitem{} Lindroos K.P., 1983, A\&AS 51, 161

\bibitem{} Lindroos K.P., 1985, A\&AS 60, 183

\bibitem{} Lindroos K.P., 1986, A\&A 156, 223

\bibitem{} Lucy L.B. and White R.L., 1980, ApJ 241, 300

\bibitem{} Mart\'{\i}n E.L., Magazz\`u A., Rebolo R., 1992, A\&A 257, 186


\bibitem{} Meyer M.R., Calvet N. and Hillenbrand L.A., 1997,
      AJ 114, 288

\bibitem{} Morrell N. and Levato H., 1991, ApJS 75, 965

\bibitem{} Neuh\"auser R., Sterzik M.F., Schmitt J.H.M.M.,
 Wichmann R. and Krautter J., 1995, A\&A 297, 391


\bibitem{} Owocki S.P., Castor J.I. and Rybicki G.B., 1988, ApJ 335, 914

\bibitem{} Panzera M.R., Tagliaferri G., Passinetti L. and Antonello E.,
 1999, A\&A 348, 161

\bibitem{} Pallavicini R., Pasquini L. and Randich S., 1992, A\&A 261, 245


\bibitem{} Quist C.F. and Lindegren L., 2000,A\&A 361, 770


\bibitem{} Rieke G.H. \& Lebofsky M.J. 1985, ApJ 288, 618

\bibitem{} Schmitt J.H.M.M, 1997, A\&A 318, 215

\bibitem{} Schmitt J.H.M.M., Zinnecker H., Cruddace R. and
 Harnden F. R. Jr, 1993, ApJ 402, L13

\bibitem{} Schmitt J.H.M.M., Fleming T.A., Giampapa M.S., 1995,
      ApJ 450, 392

\bibitem{} Schmidt--Kaler, Th., 1982, in Landolt--B\"{o}rnstein New Series,
Group VI, Vol.\ 2b, 1, eds.\ K.\ Schaifers \& H.H.\ Voigt (Springer--Verlag)

\bibitem{} Simon T. Drake S.A. and Kim P.D. 1995, PASP 107, 1034

\bibitem{} Stelzer B. and Neuh\"auser R., 2000, A\&A 361, 581

\bibitem{} Stelzer B. and Neuh\"auser R., 2001, A\&A, submitted

\bibitem{} Sterzik M. and Schmitt J.H.M.M., 1997, AJ 114, 1673

\end{thebibliography}
\end{document}